# Scandium Nitride as a Gateway III-Nitride Semiconductor for Optoelectronic Artificial Synaptic Devices

Dheemahi Rao[1,2] and Bivas Saha[1,2,3]

[1]Chemistry and Physics of Materials Unit, Jawaharlal Nehru Centre for Advanced Scientific Research, Bangalore 560064, India.
[2]International Centre for Materials Science, Jawaharlal Nehru Centre for Advanced Scientific Research, Bangalore 560064, India.
[3]School of Advanced Materials, Jawaharlal Nehru Centre for Advanced Scientific Research, Bangalore 560064, India.

Traditional computation based on von Neumann architecture is limited by the time and energy consumption due to data transfer between the storage and the processing units. The von Neumann architecture is also inefficient in solving unstructured, probabilistic, and real-time problems. To address these challenges, a new brain-inspired neuromorphic computational architecture is required. Due to absence of resistance-capacitance (RC) delay, high bandwidth and low power consumption, optoelectronic artificial synaptic devices are highly attractive. Yet stable, scalable, and complementary-metal-oxide-semiconductor (CMOS)-compatible synapses have not been demonstrated. In this work, persistence in the photoconductivity of undoped and magnesium-doped scandium nitride (ScN) is equated to the inhibitory and excitatory synaptic plasticity of the biological synapses responsible for memory and learning. Primary functionalities of a biological synapse like short-term memory (STM), long-term memory (LTM), the transition from STM-to-LTM, learning and forgetting, frequency-selective optical filtering, frequency-dependent potentiation and depression, Hebbian learning, and logic gate operations are demonstrated.

Correspondence to be addressed to bsaha@jncasr.ac.in and bivas.mat@gmail.com

Conventional von Neumann computational architecture physically partitions memory and central processing units that result in extensive time delay and high energy consumption. Moreover, the von Neumann computational architecture is inept at addressing mathematically ill-defined, nonlinear, and stochastic problems and requires complex algorithms that exert enormous resource constraints for handling big data in machine learning applications[1,2]. On the other hand, human brains are designed to solve complex spatio-temporal problems in real-time with far less energy consumption and a significant degree of fault tolerance. Their superior performance is primarily derived from the highly parallel, distributed, and event-driven computational architecture[3]. Therefore, mimicking brain operation to develop the next generations of computational architecture has gained significant attraction for over three to four decades. A human brain contains about ~ $10^{11}$ neurons that are interconnected with ~ $10^{15}$ synapses[4]. In each synapse, neurotransmitters are released from the pre-synaptic neuron in response to the action potential and flow through the synaptic cleft to the post-synaptic neuron, where the receptors detect them. Synapses are responsible for all the computation and memory in the brain. Therefore, mimicking the functionalities of the biological synaptic connection lies at the heart of the development of brain-inspired computational hardware.

Though von Neumann identified the similarities between the conventional computer architecture and the human central nervous system and tried to bring out a mathematical underpinning between the two, his untimely death in 1967 halted the substantial progress in the field for some time[5]. However, since the theoretical proposal by Carver Mead in the 1980s[6], brain-inspired neuromorphic circuits have been developed on conventional CMOS chips to emulate the synaptic functions, for example, TrueNorth chip from IBM[7], Loihi chip from Intel[8]. However, such neuromorphic integrated circuits are incredibly power-hungry, limiting their utility in cloud-based computing operations and requiring a huge number of feedback loop operations to perform simple functions such as image and voice recognition. Such large room-sized computers are also not a viable alternative to the human brain, and in recent reports, these neuromorphic circuits are also demonstrated to be prone to be biased to their training data and are found to discriminate between races and genders[9]. Therefore, though the CMOS-based neuromorphic computing technology has made great progress over the last 10-15 years, brain-inspired neuromorphic hardware technologies needs to be developed that emulates the synaptic connections in brain with significantly reduced computational power and size[10].

With the demonstration of memristor by the HP labs in 2008[11], non-volatile memory-based artificial synaptic hardware connections have gained significant attention. The history-dependent resistive switching memristor functionalities are demonstrated in several materials such as metal oxides[12], organic materials[13], perovskites[14], and 2D materials including graphene, h-BN, $MoS_2$, $WS_2$[15,16,17]. Similarly, atomic-switching memory devices with $Ag_2S$[18], ferroelectric tunnel junction devices[19], and self-forming Ag networks[20] are also utilised to demonstrate the artificial synaptic functions. However, such devices are fully electrical and suffer from RC delays from interconnects, low bandwidth, and power losses. To address these shortcomings, all photonic synaptic devices that combine neural networks' efficiency and the speed of light have been developed based on phase-change materials like gallium lanthanum oxysulfide (GLSO) microfibers[21] and GST[22]. Nevertheless, such fully optical synapses require complex wavelength conversion and expensive laser systems that are not feasible for practical applications. Therefore, optoelectrical artificial synapses that utilize persistent photoconductivity (PPC) in semiconducting materials are proposed that offer bidirectional conversion between the electrical and optical signals[23]. Such photoelectric synapses exhibit wide bandwidth, lower RC delay and power consumption. They can also integrate visual sensing, signal processing, and memorising that closely emulates the human biological visual cortex system.

Optoelectronic artificial synaptic devices made up of metal oxides[24], halide perovskites[25], carbon nanotube heterostructures[26], organic heterojunctions[27], nanocrystals[28], and 2D-MoS$_2$[29] have successfully mimicked the primary biological neural functions like STM, LTM, paired-pulse facilitation, and spike timing dependent plasticity (STDP). For artificial neural network application, tasks such as learning-forgetting[24], classical conditioning[24,27], latent inhibition[24], aversion[28], logic functions[27,28], dynamic filtering[30,29], and pattern sensing and memorizing[27] are demonstrated in some of these optoelectronic artificial synaptic devices. While such proof of concept demonstration has invigorated the optoelectronic artificial synapse research field, for practical device implementations where the artificial synapse-based neural circuits will have to be integrated in the CMOS chips (at least in the initial states of development), stable, scalable, and robust materials are required. Wurtzite III-nitride semiconductors, such as GaN, InN, and AlN are stable, CMOS processable, scalable and attractive for a wide range of optoelectronic applications. However, their fast carrier recombination rate and little PPC are not commensurate for developing optoelectronic synapses[31,32].

Scandium nitride (ScN) is an emerging group-III (B) semiconducting nitride that crystallises in the rocksalt structure with octahedral coordination and is stable in ambient conditions with high mechanical hardness[33]. ScN has found significant interest for its thermoelectric properties[34], as a substrate for defect-free GaN growth[35], and as a semiconductor in metal/semiconductor superlattice development[36]. ScN exhibits an indirect bandgap of ~ 0.9 eV and a direct bandgap of ~ 2.2 eV[37,38]. ScN thin films are highly degenerate due to the presence of oxygen as impurity (O$_N$) and nitrogen vacancies (V$_N$) exhibiting a carrier concentration in ~ (1-4) × $10^{20}$ cm$^{-3}$ range[39,40]. ScN thin films exhibit mobility between 20-120 cm$^2$/Vs at room temperature depending on the growth conditions[41,42,43]. Mg-doping is found to be effective in reducing the high electron concentration, and *p*-type ScN thin films have been demonstrated that also exhibit high thermoelectric power factor[44,45].

This work utilises persistent negative photoconductivity (PNPC) in intentionally undoped ScN and persistent positive photoconductivity (PPPC) in Mg-doped ScN thin films to develop inhibitory and excitatory optoelectronic artificial synapses, respectively. These artificial synapses exhibit STM and LTM, transitions from STM-to-LTM, learning, frequency-controlled paired-pulse facilitation (PPF) and paired-pulse depression (PPD), dynamic filtering, symmetric spike-timimg dependent plasticity (STDP) and logic gate operations. Temperature-dependent photoconductivity and photo-Hall measurements are further performed to explain the underlying device operation mechanism.

Intentionally undoped ScN and Mg-doped ScN thin films are deposited on (001) MgO substrates with reactive magnetron sputtering in an ultra-high vacuum chamber. Without any intentional doping, the ScN film has electron concentration of 2×$10^{20}$ cm$^{-3}$ and 63 cm$^2$/Vs mobility. O$_N$ and V$_N$ act as unintentional n-type dopants in ScN making it a degenerate semiconductor. With Mg(hole)-doping, the electron concentration is reduced to ~2×$10^{18}$ cm$^{-3}$, and at higher Mg concentration holes become the majority carriers. Two-terminal devices are fabricated with indium (In) Ohmic contacts. The methods and supplementary section includes details on the growth, characterisation, device fabrication, and measurements.

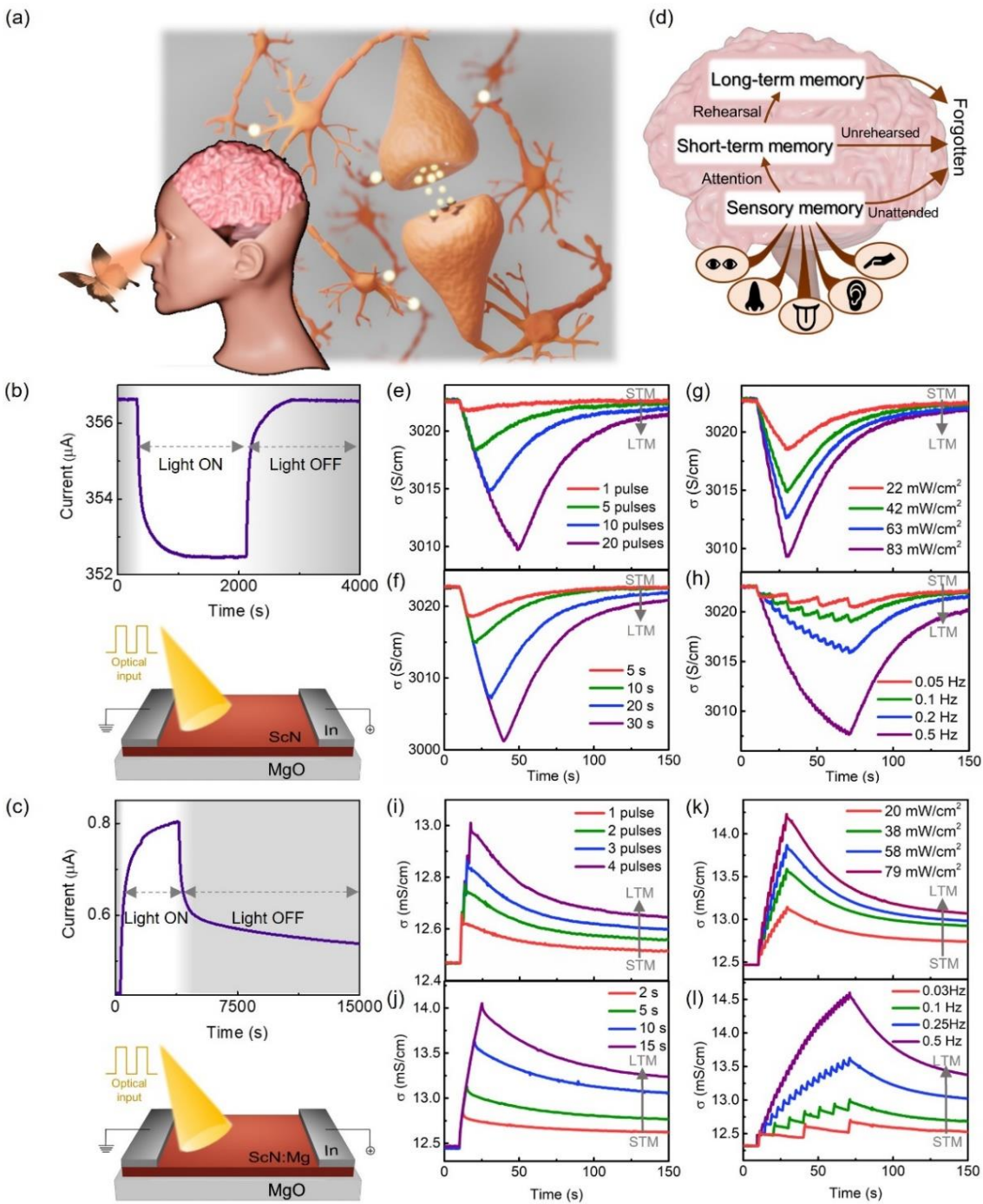

Figure 1: (a) Schematic of human visual system and the neural synapse. (b) Negative photoconductivity in ScN thin film measured in the device geometry shown below. At the onset of illumination, the current starts to decrease gradually and saturates within 30 minutes. After the light is turned off, the current takes about 15 minutes to return to its initial value. (c) Positive photoconductivity in Mg-doped ScN thin film is measured in the device geometry shown

below. At the onset of illumination, the current starts to increase gradually and does not saturate till 60 minutes. After turning off the light, the current takes longer than 3 hours to return to its initial value. (d) Atkinson–Shiffrin memory model proposing three main stages of memory in human brain. (e)-(h) Transition from STM to LTM in inhibitory ScN synapse as a function of number, duration, intensity and frequency of the optical pulses. (i)-(l) Transition from STM to LTM in excitatory Mg-doped ScN synapse as a function of the number, duration, intensity and frequency of the optical pulses.

The external stimuli sensed by the sensory organs generate an action potential that will release the neurotransmitters from the pre-synapse (Fig 1(a)). Depending on their role, these neurotransmitters will either increase (excitatory) or decrease (inhibitory) the likelihood of depolarisation and generation of the action potential at the post-synapse. Figure 1(b) shows that undoped ScN exhibits negative photoconductivity, where the current decreases on illumination. After switching off the light, the current takes several minutes to return to its initial value. On the other hand, the current increases on illumination in Mg-doped ScN, as commonly seen in many semiconductors (Fig. 1(c)). This positive photocurrent persists for several hours to several days, depending on the Mg concentration. Thus, PNPC in degenerate ScN and PPPC in Mg-doped ScN can be used as inhibitory and excitatory actions of a synapse, respectively with the persistence in photoconductivity equated to the synaptic plasticity. Synaptic plasticity plays a vital role in creating memory. The higher the plasticity, the longer is the memory retention.

According to the Atkinson–Shiffrin model, memory development in the human brain has three main stages (as shown in Fig. 1(d))[46]. First, the sensory organs sense the external stimuli and form sensory memory at the superficial layers of the brain. Paying attention to this input will transform it into STM lasting only for a few seconds to minutes. On rehearsal or revision, this data is stored as LTM that will be retained for several days, years, or a person's entire lifetime. This aspect of memory formation is successfully imitated in both the inhibitory and excitatory ScN-based artificial synapses. The effect of rehearsal is achieved by increasing the strength of the optical pulses by increasing their number, power, duration, and frequency. The STM, LTM, and transition from STM-to-LTM is executed in inhibitory (Fig. 1(e-h)) and excitatory (Fig. 1(i-l)) synapses with optical pulses. When the strength of the stimuli is low, the persistence in NPC and PPC is short (a few seconds), implying STM. With an increase in the stimuli strength in the form of more pulses, higher intensity, larger exposure time and higher frequency, the persistence of photocurrent in both inhibitory and excitatory synapse increases that can be equated to their LTM. A clear transition from the STM-to-LTM is observed on all occasions, equivalent to rehearsing the information to remember it for an extended period.

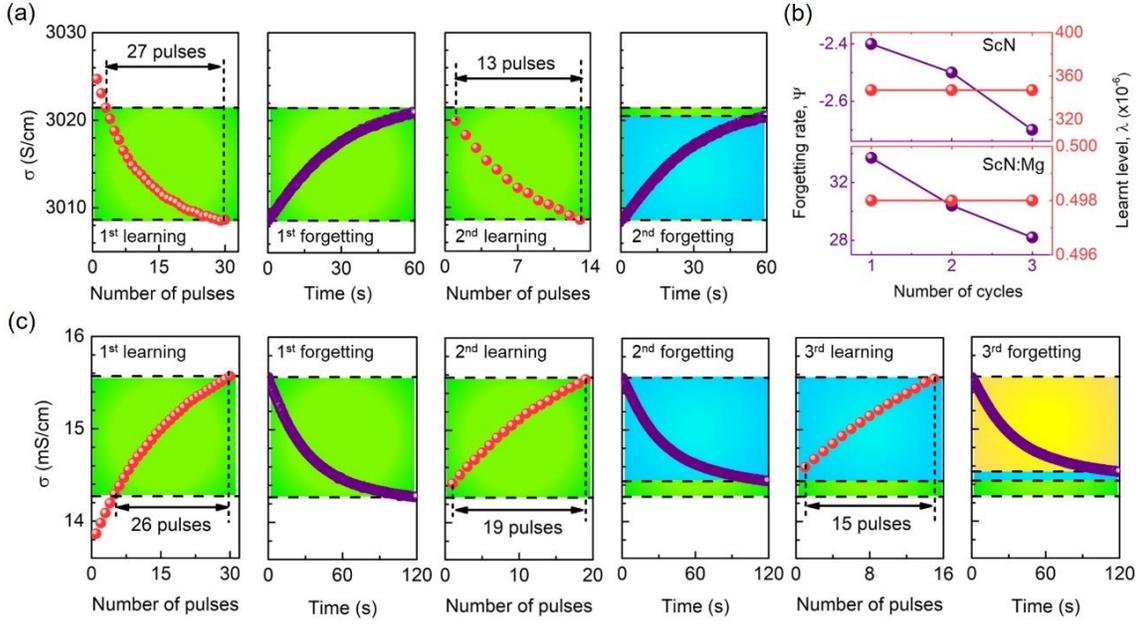

Figure 2: (a) Learning and forgetting cycles in ScN inhibitory artificial synapse. (b) Wickelgren's power law fitting clearly indicates the decrease in the forgetting rate with increase in number of learning cycles. (c) Learning and forgetting cycles in Mg doped ScN excitatory artificial synapse.

Along with memory, learning is also an essential function of the brain and is the foundation for building brain-like computers. The change in conductivity due to the optical pulses is considered learning of the device, and restoration to the initial level is considered forgetting. In both inhibitory and excitatory ScN synaptic devices, the number of pulses required to reach a certain conductivity threshold reduces in consecutive cycles. Also, the rate of decay reduces with the number of cycles. The behaviour of these devices follows Ebbinghaus's learning and forgetting curve[47]. According to this model, first-time learning takes much effort, and the learned information is forgotten over a while. However, re-learning takes less effort, and the learnt information is remembered for a longer time. In figure 2(a), ScN inhibitory synapses initially take 27 pulses for learning, and the learned information is forgotten in a minute. The second time learning requires only 13 pulses, and forgetting is slower than in the previous cycle. Similarly, in Mg-doped ScN excitatory synapses (Fig 2(c)), the number of pulses required in learning cycles reduces from 26 to 15 in three consecutive cycles, along with the decrease in the extent of forgetting. The learning-forgetting behaviour in ScN can be better analysed with Wickelgren's power law[48]. As per this law, the transient current (I) after the removal of stimuli (forgetting curves) can be fitted as a function of time (t) with the following equation:

$$I = \lambda(1+\beta t)^{-\psi}$$

where $\lambda$ is the learned memory state, $\beta$ is the scaling parameter, and $\psi$ is the forgetting rate parameter. The $\lambda$ and $\psi$ obtained from the fitting of all forgetting curves are presented in figure 2(b). The forgetting rate parameter is negative because of the negative photoresponse of undoped ScN. From the same state of learnt memory, the forgetting rate parameter decreases with the increasing number of learning cycles in both excitatory and inhibitory artificial synapses. This clearly presents the learning ability of the ScN artificial optoelectronic synapses.

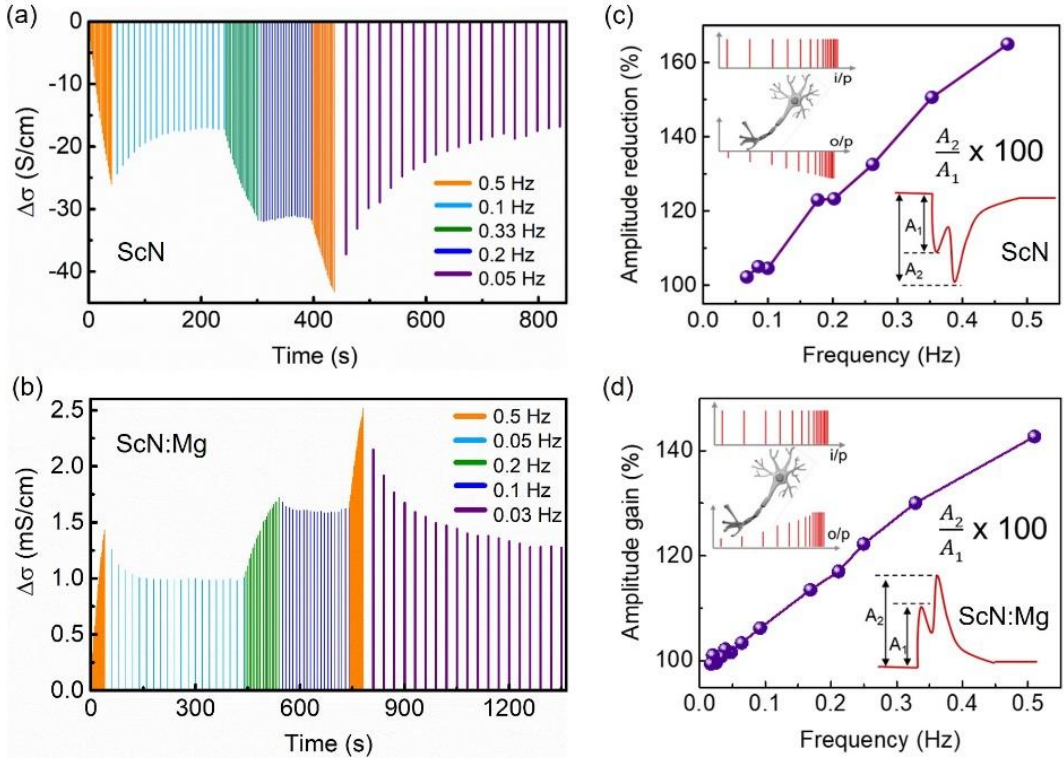

Figure 3 : Frequency dependent PPF and PPD in (a) ScN and (b) Mg-doped ScN artificial synapse. (c) The amplitude of current reduction increases at higher frequencies in inhibitory synapse functioning like a low pass filter. (d) The amplitude of gain increases at higher frequency, thus acting like high pass filter.

Post-synaptic current can be modulated as a function of the temporal pattern of the pre-synaptic signals in the brain. The frequency of the train of optical pulses arriving at the pre-synapse in the ScN synaptic devices alters the post-synaptic current depending on the memory in the synapse stored due to preceding signals. In the train of optical pulses, the successive pulses can either increase (facilitate) or decrease (depress) the synaptic strength depending on the time interval between them. This well-known frequency dependence of PPF and PPD is exhibited by both inhibitory and excitatory ScN artificial synapses, as shown in Fig. 3(a) and Fig. 3(b), respectively. Here, on applying a set of 20 optical pulses at different frequencies switching between PPF and PPD is achieved.

In addition, biological synapses also act as temporal filters for information processing based on stimulation frequency. Such a process is emulated in ScN artificial synaptic devices by measuring the post-synaptic current as a function of frequency of the optical stimuli. The frequency-dependent amplitude reduction (amplitude gain) of post-synaptic current in the inhibitory (excitatory) synapse is calculated as $|A_2/A_1| \times 100$. In ScN synapse, low-frequency signals are less inhibited, and hence their transmission to post-synaptic neurons is more probable (Fig. 3(c)). In contrast, higher frequency signals have high amplitude gain in Mg-doped ScN and have higher transmittance probability(Fig. 3(d)). Thus, ScN inhibitory synapse and Mg-doped ScN excitatory synapse act as low pass and high pass temporal filters, respectively.

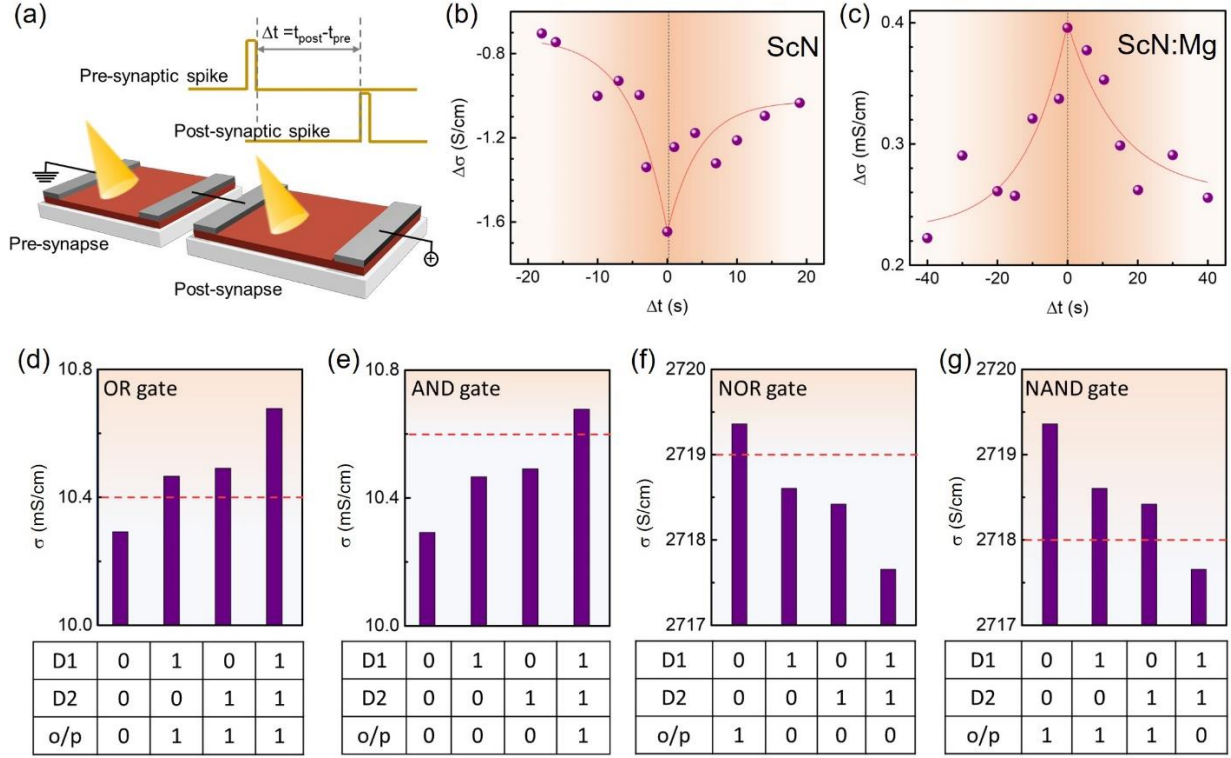

Figure 4: (a) Schematic of measurements done with two devices connected in series. Symmetric STDP curve representing Hebbian learning in (b) inhibitory ScN synapse and (c) excitatory Mg-doped ScN synapse. (d) OR and (e) AND logic operations demonstrated with two Mg-doped ScN devices in series. (f) NOR and (g) NAND logic operations demonstrated with two ScN devices in series.

While the above demonstrations consider stimulation of only pre-synapse, synaptic plasticity can also be varied by correlated stimulation of both pre- and post- synapses. The simple theory explaining this process in biological neurons is Hebb's postulate that states, "neurons that fire together wire together"[49]. This concept can be emulated in ScN artificial synapses using two devices in series, one pre-synapse and the other post-synapse (Fig 4(a)). When the optical stimulus is incident on both the synapses simultaneously ($\Delta t=0$), the change in the net conductivity is maximum, indicating a strong synaptic connection. As the time between the pre-synaptic and post-synaptic stimulus increases, the net change in synaptic conductivity reduces, as shown in figure 4(b) and figure 4(c). This symmetric curve representing the Hebbian law is one of the types of spike timing-dependent plasticity (STDP), an essential phenomenon in determining the synaptic strength in biological neurons.

Further, information processing requires synapses to perform logic operations. The logic gates in the ScN artificial synapses are demonstrated with two devices connected in series (Fig. 4(a)). If the optical pulse is incident on the device (D1 or D2), it is denoted as 1, else 0. The post-synaptic conductivity resulting from the optical stimuli is considered 1 if it is higher than a set threshold, else considered 0. The net conductivity is low when two Mg-doped ScN devices are connected in series without any input optical pulse. The optical stimuli on either or both the devices will increase the net conductivity due to their positive photoconducting nature. By setting a suitable threshold (as red dashed lines in Fig. 4(d) and 4(e)), OR and AND gate operations are exhibited utilising the excitatory synapses. Similarly, with inhibitory ScN synapses in series, NOR and NAND gate operations are demonstrated employing their negative photoconductivity (Fig. 4(f) and Fig. 4(g)).

Since neuromorphic computers are required to exhibit high degrees of energy efficiency, measuring the energy consumption of the ScN artificial synapses is critical. The power density of the ScN excitatory and inhibitory synapses are calculated using the following equation:

$$\frac{dW}{A.dt} = V \times \Delta I$$

where, $W$ is the total power, $dt$ is the pulse width, and A is the device area. V and $\Delta I$ are the applied voltage across the electrodes (read voltage) and the photocurrent, respectively. Calculations show that the power density of inhibitory synapse is 0.13 nW/mm$^2$ at a read voltage of 20 mV, and that of the excitatory synapse is 0.65 nW/mm$^2$ at 1V read voltage. A relatively higher read voltage is used for excitatory synapse because of its higher resistivity. Though the previous works on optoelectronic synapses have reported the absolute energy consumption, their corresponding power density is found to be similar to the ScN synapses (more details in supplementary material).

After demonstrating synaptic functionalities, we address the positive and negative photoconduction mechanisms in ScN that is vital for emulating excitatory and inhibitory synaptic activities. In Mg-doped ScN, the Fermi level is in the bandgap (Fig. 5(a)). The positive photoconductivity results from the light-induced electron-hole pair generation, and collection of the photogenerated carriers at the electrodes, similar to most conventional semiconductors' photoresponse. However, in pristine ScN, the Fermi level resides inside the conduction band by ~ 100-250 meV due to the unintentional oxygen (*n*-type) incorporation (Fig. 5(b))[45]. The photogenerated electrons are excited to the higher states in the conduction band (above $E_F$). The holes left behind in the valence band are captured by the trap states in the bandgap ($V_N$ in ScN)[38]. These captured holes act as charged scattering centres for photogenerated electrons, resulting in mobility reduction. Since the carrier concentration is already high, the net increase in the carrier density due to illumination is expected to be very small. This leads to a net decrease in conductivity (neµ) on illumination seen as negative photoconductivity. Photo-Hall measurement of the pristine ScN is performed to verify this mechanism. Results show that with light ON, the carrier concentration (n) (Fig. 5(e)) remains almost unchanged, while the mobility (µ) (Fig. 5(f)) decreases. The product of carrier concentration and mobility, defining the conductivity (neµ) (Fig. 5(g)) follows the same trend as the photoconductivity (Fig. 5(h)). It is important to note that the same model also successfully explained the negative photoconduction in degenerate InN[50]. Temperature-dependent photoresponse of pristine and Mg-doped ScN are measured in the 80-400 K temperature range (Fig. 5(c) and 5(d)). The nature of photoresponse remains unaltered, while the magnitude of photoconductivity change increases at lower temperatures. A detailed analysis of the temperature-dependent growth and decay of photoconductivity is presented in the supplementary material.

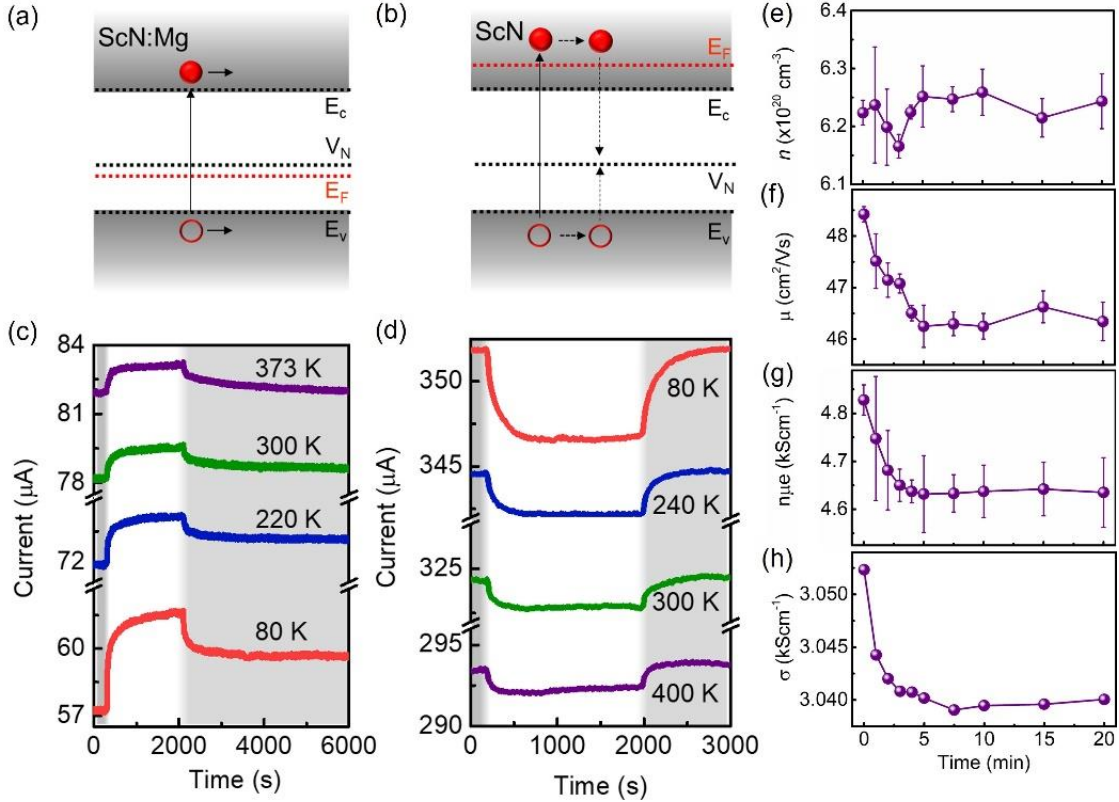

Figure 5: (a) Band diagram of Mg-doped ScN where photogenerated carriers enhance the current in the device. (b) Band diagram explaining the negative photoconductivity in pristine ScN where the photogenerated electrons are scattered by the charged scatterers formed at the sub-bandgap state ($V_N$). The nature of photoconductivity of (c) pristine ScN and (d) Mg-doped ScN measured at various temperatures reveals the robustness of this phenomenon. (e) carrier concentration and (f) mobility of ScN measured with light being incident on the sample. The carrier concentration change is within the error bar while the mobility clearly decreases. (g)The product nμ decreases, indicating a lowering of conductivity (nμe). (h) The negative photoconductivity observed in pristine ScN matches well with (g) obtained from the photo-hall effect.

In conclusion, inhibitory and excitatory synaptic functionalities are demonstrated in CMOS-compatible group-III semiconducting ScN thin films by controlling carrier concentration with Mg-(hole) doping. The persistence in photoconductivity is equated to memory in such optoelectronic synapse. STM, LTM, and transition from STM-to-LTM are demonstrated as a function of the strength of the input optical pulse tuned by its duration, number, power, and frequency. Important synaptic activities like Ebbinghaus learning and forgetting cycles, frequency-dependent PPF and PPD, and temporal filtering of signals are emulated. The Hebbian learning and logic gate operations are successfully implemented with two synaptic devices in series. The negative and positive photoresponse of degenerate ScN and Mg-(hole)-doped ScN depends on the position of the Fermi level. This work paves the way for neuromorphic computer hardware development with a CMOS compatible, high-temperature stable III-nitride semiconductor.

**Online Methods:**

Single-crystalline scandium nitride thin films are deposited on 1 cm × 1 cm single-side polished MgO (001) substrates with reactive DC magnetron sputtering. The substrates are ultrasonicated in acetone and methanol before loading into the sputtering ultra-high vacuum chamber with $1\times10^{-9}$ Torr base pressure. An Ar:$N_2$ gas mixture with ratio of 9:2 is used, and the substrate temperature is maintained at 800°C during all depositions. Sc target DC power is set to 100 W for pure ScN deposition, while it is varied in relation to Mg power during deposition of Mg doped ScN to achieve different doping levels. The Mg-doped sample presented here is deposited at 125 W Sc and 6 W Mg power.

Indium wire (Alfa Aesar, 99.999% purity) is pressed onto the films to make Ohmic contacts. The current is measured using Kiethley 2450 with a read voltage of 0.02 V and 1 V for ScN and ScN:Mg, respectively. The entire spectrum of the Newport Oriel Xe- arc lamp covering 250-2400 nm is incident on the sample without any filters or monochromator. All the photocurrent measurements are performed in vacuum environment at ~ $5\times10^{-6}$ torr. Janis cryostat is used for temperature-dependent photoconductivity measurement. Apart from intensity-dependent measurements, the photocurrent measurements are performed with 40 mW/cm$^2$ optical intensity falling on the samples. Optical pulses of 1 second are used for all the measurements except for pulse duration-dependent measurement.

*Trans. Biometrics, Behav. Identity Sci.* **3**, 101–111 (2021).

10. Schuman, C. D. *et al.* A Survey of Neuromorphic Computing and Neural Networks in Hardware. 1–88 (2017).

11. Strukov, D. B., Snider, G. S., Stewart, D. R. & Williams, R. S. The missing memristor found. *Nature* **453**, 80–83 (2008).

12. Deswal, S., Kumar, A. & Kumar, A. NbOx based memristor as artificial synapse emulating short term plasticity. *AIP Adv.* **9**, 6–11 (2019).

13. Goswami, S. *et al.* Decision trees within a molecular memristor. *Nature* **597**, 51–56 (2021).

14. Xiao, Z. & Huang, J. Energy-Efficient Hybrid Perovskite Memristors and Synaptic Devices. *Adv. Electron. Mater.* **2**, 1–8 (2016).

15. Seo, S. *et al.* Artificial optic-neural synapse for colored and color-mixed pattern recognition. *Nat. Commun.* **9**, 1–8 (2018).

16. Huh, W., Lee, D. & Lee, C. H. Memristors Based on 2D Materials as an Artificial Synapse for Neuromorphic Electronics. *Adv. Mater.* **32**, 1–16 (2020).

17. Arnold, A. J. *et al.* Mimicking Neurotransmitter Release in Chemical Synapses via Hysteresis Engineering in MoS2 Transistors. *ACS Nano* **11**, 3110–3118 (2017).

18. Ohno, T. *et al.* Short-term plasticity and long-term potentiation mimicked in single inorganic synapses. *Nat. Mater.* **10**, 591–595 (2011).

19. Li, J. *et al.* Reproducible Ultrathin Ferroelectric Domain Switching for High-Performance Neuromorphic Computing. *Adv. Mater.* **32**, 1–9 (2020).

20. Bannur, B. & Kulkarni, G. U. On synapse intelligence emulated in a self-formed artificial synaptic network. *Mater. Horizons* **7**, 2970–2977 (2020).

21. Gholipour, B. *et al.* Amorphous Metal-Sulphide Microfibers Enable Photonic Synapses for Brain-Like Computing. *Adv. Opt. Mater.* **3**, 635–641 (2015).

22. Feldmann, J., Youngblood, N., Wright, C. D., Bhaskaran, H. & Pernice, W. H. P. All-optical spiking neurosynaptic networks with self-learning capabilities. *Nature* **569**, 208–214 (2019).

23. Wang, Y. *et al.* Optoelectronic Synaptic Devices for Neuromorphic Computing. *Adv. Intell. Syst.* **3**, 2000099 (2021).

24. Ng, S. E., Yang, J., John, R. A. & Mathews, N. Adaptive Latent Inhibition in Associatively Responsive Optoelectronic Synapse. *Adv. Funct. Mater.* **31**, 1–12 (2021).

25. Wang, Y. *et al.* Photonic Synapses Based on Inorganic Perovskite Quantum Dots for Neuromorphic Computing. *Adv. Mater.* **30**, 1–9 (2018).

26. Pilarczyk, K. *et al.* Synaptic Behavior in an Optoelectronic Device Based on Semiconductor-Nanotube Hybrid. *Adv. Electron. Mater.* **2**, 2–5 (2016).

**Acknowledgments:** DR and BS acknowledge International Centre for Materials Science (ICMS) and Sheikh Saqr Laboratory (SSL) in JNCASR for support. B.S. acknowledges the Science and Engineering Research Board (SERB) of the Government of India, Start-Up Research Grant No. SRG/2019/000613 for financial support. DR thanks JNCASR for the fellowship. DR acknowledges SAMat research facilities. DR acknowledges Dr.Shashidhara Acharya and Krishna Chand Maurya for assistance in experimental work.

**Author contributions:** DR and BS conceived this project. DR deposited the thin films and performed materials characterisation, device fabrication, and transport measurement experiments. DR and BS analysed the results and prepared the manuscript.

**Additional information:** Supplementary information is available in the online version of the paper. Reprints and permissions information is available online at   .

Correspondence and requests for materials should be addressed to BS.

**Competing interests:** The authors declare no competing interests

# Supplementary Information

## Scandium Nitride as a Gateway III-Nitride Semiconductor for Optoelectronic Artificial Synaptic Devices

Dheemahi Rao[1,2] and Bivas Saha[1,2,3]

[1]Chemistry and Physics of Materials Unit, Jawaharlal Nehru Centre for Advanced Scientific Research, Bangalore 560064, India.
[2]International Centre for Materials Science, Jawaharlal Nehru Centre for Advanced Scientific Research, Bangalore 560064, India.
[3]School of Advanced Materials, Jawaharlal Nehru Centre for Advanced Scientific Research, Bangalore 560064, India.

1. **Structural Characterization**

ScN thin films exhibit rocksalt crystal structure and grow as epitaxial nominally single-crystalline with (001) orientations on (001) MgO substrates. High-resolution x-ray diffractogram (HRXRD) performed with a Bruker D8 Discover show two peaks at 39.99° and 42.91° that correspond to the (002) planes of ScN and MgO, respectively (Fig. S1(a)). From the x-ray diffraction peak positions, a lattice constant of 4.50 Å is measured for ScN. Mg-hole doping does not shift the (002) peak position but increases the FWHM of the rocking curve ($\omega$-scan) from 0.74° to 1.03° (see inset in Fig. S1(a)) suggesting a slight degradation in the crystalline quality due to Mg-incorporation.

The optical bandgap of the ScN thin films is measured with Perkin Elmer lambda750 spectrometer. The transmission and absorbance of the thin films were recorded in the 250-850 nm wavelength range and the absorption co-efficient($\alpha$) is calculated using the formula below.

$$\alpha = \frac{1}{thickeness} \times \ln\left(\frac{(1-Reflectance)}{Transmittance}\right)$$

From the Tauc's plot ( $(\alpha h\nu)^2$ Vs. $h\nu$) the band gap of both pristine ScN and Mg-doped ScN is extracted to be 2.26 eV (Fig. S1(b)). Mg-doping does not alter the bandgap or create sub bandgap states as previously reported[1,2].

Room temperature electrical properties are measured using Ecopia HMS3000 system. The carrier concentration, mobility and resistivity of the samples discussed in this paper are listed in the table S1. Mg-doped ScN-2 is used only for temperature-dependent photoconductivity measurement and for all other measurements, Mg doped ScN-1 is used.

The resistivity of pristine ScN increases with increasing temperature. This suggests that the pristine ScN is a degenerate semiconductor with Fermi-level inside the conduction band. The Fermi level is lowered into the bandgap on Mg-hole doping. As a result, the resistivity reduces at higher temperatures as in normal semiconductors ( Fig. S1(c)).

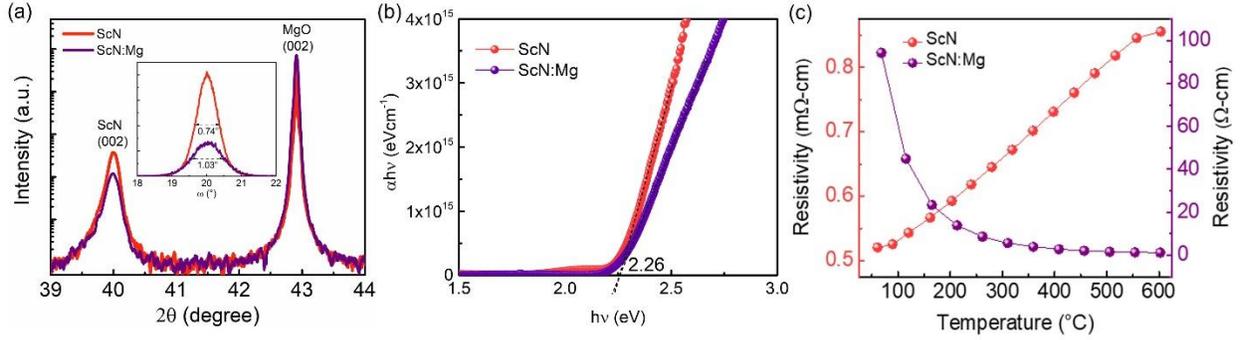

Figure S1: (a) HRXRD of both undoped and doped ScN have peak at 39.99° indicating epitaxial nominally single-crystalline growth of the films. The rocking curve (ω-scan) in the inset reveals higher FWHM of the Mg-doped ScN in comparison to undoped ScN. (b) The Tauc plot shows that the bandgap of both pristine and the Mg-doped ScN is 2.26 eV. (c) Temperature-dependence of the resistivity of the films suggests that undoped ScN is a degenerate semiconductor, while Mg-doped ScN is a conventional semiconductor whose resistivity decreases at higher temperatures.

Table S1: Room temperature electrical properties of the samples reported in this work.

| Sample | Carrier concentration (cm$^{-3}$) | Mobility (cm$^2$/Vs) | Resistivity (ohm-cm) | Thickness (nm) |
|---|---|---|---|---|
| $n$-type ScN | $3 \times 10^{20}$ electrons | 67 | 0.0003 | 250 |
| Mg doped ScN-1 | $2 \times 10^{17}$ holes | 0.2 | 50 | 200 |
| Mg doped ScN-2 | $5 \times 10^{18}$ holes | 9.7 | 0.127 | 220 |

The plan-view FESEM images (captured with FEI Inspect F50) show the smooth surface of ScN films with some square or pyramidal plateaus occurring due to defects as mentioned in earlier reports[3] (Fig. S2(a) and S2(b)). The surface roughness is found to be 1 nm and 0.6 nm for the undoped and Mg-doped ScN films, respectively, as measured from atomic force micrographs (Fig. S2(c) and S2(d)).

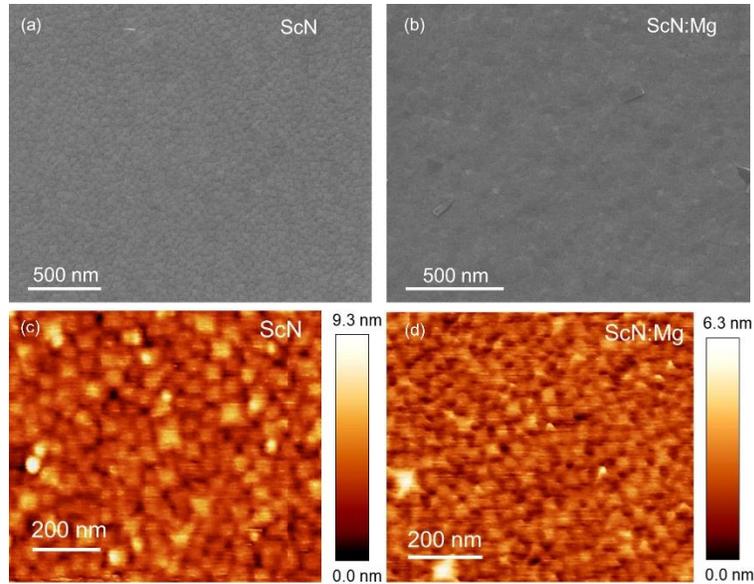

Figure S2: Plan-view FESEM image of (a) ScN and (b) Mg-doped ScN revealing the smooth surface of the films with a few square/triangular (tilted cube) mounds arising due to the defects. Atomic probe micrographs of (c) ScN and (d) Mg-doped ScN showing the smooth squared grains on the surface of the films.

## 2. Photocurrent Measurements

Indium is used as Ohmic contacts on both intentionally undoped ScN and Mg-doped ScN films. The linear I-V characteristic across the two lateral In contacts verifies the formation of Ohmic contact (Fig. S3). Hence, the role of metal/semiconductor interface barriers in the observed photoresponse can be neglected. The undoped and Mg-doped ScN devices dimensions are ~ 7.2 mm × 1.7 mm and 6.1 mm × 2.3 mm, respectively.

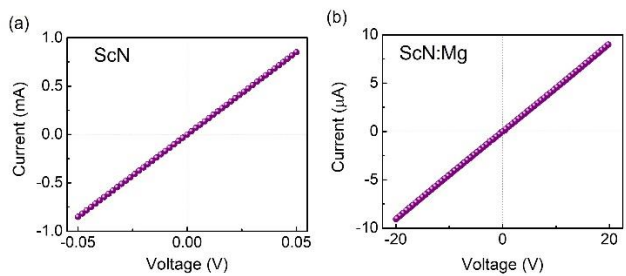

Figure S3: Indium metal makes ohmic contact with both (a)ScN and (b)Mg-doped ScN.

## 3. Photo-Hall Measurement

To study the change in the carrier concentration and mobility due to illumination, the Hall measurement were performed with light incident on the samples (Fig. S4(a)). The white light from Xe-arc lamp is turned ON at t=0 s and the change in electrical properties is noted as a function of time. For degenerate ScN with ~ $2 \times 10^{20}$ cm$^{-3}$ electrons, the mobility reduces significantly without much change in carrier concentration resulting in the reduction in conductivity (Fig. 5 in main manuscript). In Mg-doped ScN with Fermi level inside the bandgap, the photo generated holes leads to increased carrier concentration, and the mobility remains almost same (Fig. S4(b) and S4(c)). The product (pμe) shows that the conductivity increases on illumination explaining the positive photoconductivity (Fig. S4(d)).

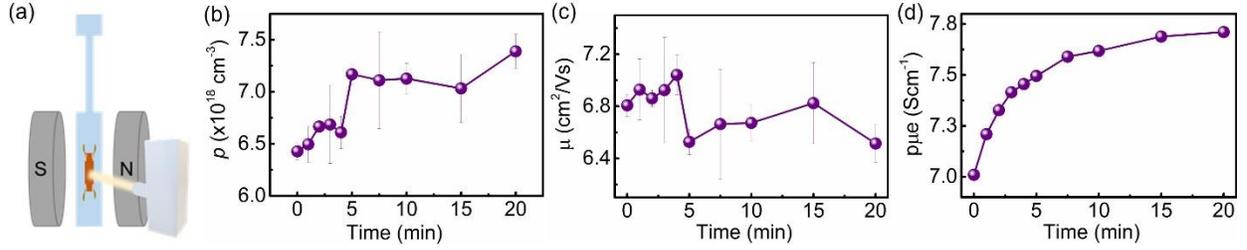

Figure S4: (a) Schematic of the photo-hall measurement setup. (b), (c) and (d) are the variation of hole concentration, mobility and conductivity in Mg-doped ScN with light. The photo-generated holes increase the conductivity resulting in positive photoconductivity.

## 4. Mathematical fittings of Photoconductivity

To gain more insights into the processes involved in the photo-response, the transient photocurrent curves are fitted with exponential functions. The negative photoconducting response of pristine ScN and the regeneration after turning off the light are fitted with double exponential functions (equations 1 and 2). A representative fitting for photocurrent measured at 80 K is presented in figure S5(a). During the recovery of the photocurrent, the time constants (τ) of the two processes are an order of magnitude apart (Fig. S6(a)). The faster process has $\tau_1$ ~ 25 s and for the slower one $\tau_2$ ~ 200 s at 80 K. Both the time constants decreases with an increase in temperature.

The photo current enhancement in Mg-doped ScN follows double exponential function (equation 3) with two time constants in the range of 10s and 400s. After turning off the light, the photocurrent decays as the sum of three exponentials (equation 4). A representative fitting of the photo response measured at 100 K is presented in figure S6(b). In most previous studies, the persistence photoconductivity is seen to follow the stretched exponential behaviour (equation 5). But three exponential decay functions fit the photocurrent decay curves of Mg doped ScN device studied here better, in comparison to the stretched exponential function. The largest among the three time constants ($\tau_3$), is in the range of $10^5$ s and decreases at higher temperature (Fig. S6(b)). This component is responsible for the persistent photocurrent for more than 30 hours, as seen in the experiments. The persistence in photocurrent is seen to be higher in samples with higher Mg concentrations.

The Arrhenius fit for $\tau_3$ yields an activation energy of ~ 30 meV (Fig. S6(c)). This energy being slightly higher than the room temperature energy, is responsible for the persistent photoconductivity at room temperature. Complex defect centres like DX/AX – centres are found to have such neutralization energies that result in slow photocurrent decay [4]. At 373 K, the photocurrent decays significantly faster than in room temperature as the activation energy required for recombination is available at higher temperature. The decay curve also follows stretched exponential function indicating the change in underlying processes at 373 K (Fig. S5(c)). An increase in recombination rate at 373K hints at the process being thermally activated.

For pure ScN:    $I = I_o + [A_1 \exp(-(t-t_o)/\tau_1)] + [A_2 \exp(-(t-t_o)/\tau_2)]$    {1}

$I = I_o + [A_1\{1-\exp(-(t-t_o)/\tau_1)\}] + [A_2\{1-\exp(-(t-t_o)/\tau_2)\}]$    {2}

For Mg doped ScN:    $I = I_o + [A_1\{1- \exp(-(t-t_o)/\tau_1)\}] + [A_2\{1- \exp(-(t-t_o)/\tau_2)\}]$    {3}

$I = I_o + [A_1 \exp(-(t-t_o)/\tau_1)] + [A_2 \exp(-(t-t_o)/\tau_2)] + [A_3 \exp(-(t-t_o)/\tau_3)]$    {4}

Stretched exponential:    $I = I_o + A \exp(-((t-t_o)/\tau)^\beta)$    {5}

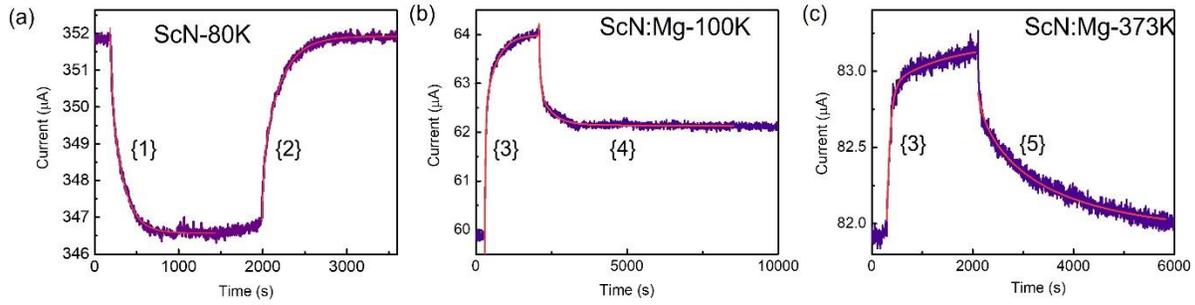

Figure S5: (a) Double exponential function given in equation 1 and 2 fits the photo response in ScN at 80K very well. (b) The photocarrier generation in Mg-doped ScN is well fitted with double exponential equation 3, and the decay follows triple exponential function (equation 4) at 100K. (c) At 373K, the photocurrent decay in Mg-doped ScN follows stretched exponential function (equation 5).

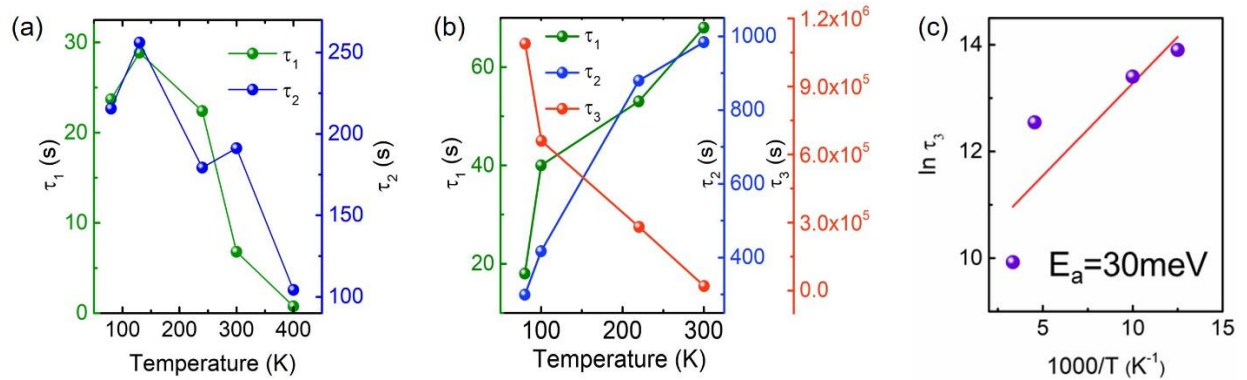

Figure S6: Temperature dependence of the time constants obtained from exponential fittings of the persistent photocurrent in (a) ScN and (b) Mg doped ScN. (c) Arrhenius fit of the time constant $\tau_3$ of Mg-doped ScN which is responsible for long-term memory yields activation energy of 30 meV.

## 5. Effects of Deposition Techniques and Substrates on the Photo response

To verify the presence of negative photo response in ScN deposited with other growth methods, we conducted ScN growth with a plasma-assisted molecular beam epitaxy (MBE). See reference [5] for details on the growth and characterisation. The negative photo response of intentionally undoped ScN is also present in films deposited with MBE (Fig. S7(a)). Hence, the negative photo response in ScN is universal and do not depend on the film growth technique.

In order to check if the photo response of ScN depends on the substrate on which the film is deposited, undoped ScN and Mg-doped ScN thin films are also deposited on sapphire substrates. On illumination, pristine ScN exhibits negative photoconductivity, while Mg-doped ScN exhibits positive photoconductivity as in figure S7(b) and S7(c). Thus, the photo response is not specific to ScN/MgO device.

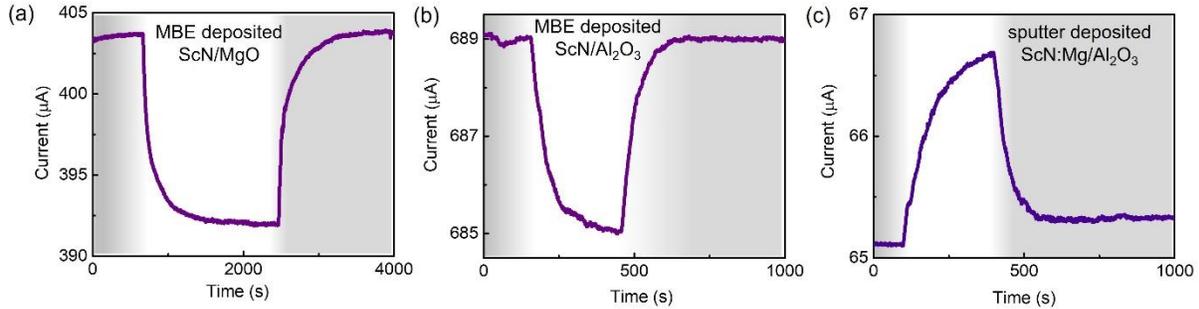

Figure S7: Photoresponse of (a) undoped ScN deposited on (001) MgO substrates with plasma-assisted MBE, (b) MBE-deposited ScN on (0001) sapphire substrate, and (c) Mg-doped ScN deposited on (0001) sapphire with magnetron sputtering is presented.

The amount of photo response in ScN and in Mg-doped ScN is found to be higher in vacuum than in ambient condition (Fig. S8(a) and S8(b)). This leads to longer persistence in photocurrent in vacuum. Bandpass filters are also used to find the spectral contribution of the photo-response. It is seen that the visible and UV light contribute to most of the photo-response. Yet, there is some response for wavelengths higher than 1500 nm as well. (Fig. S8(c) and S8(d)).

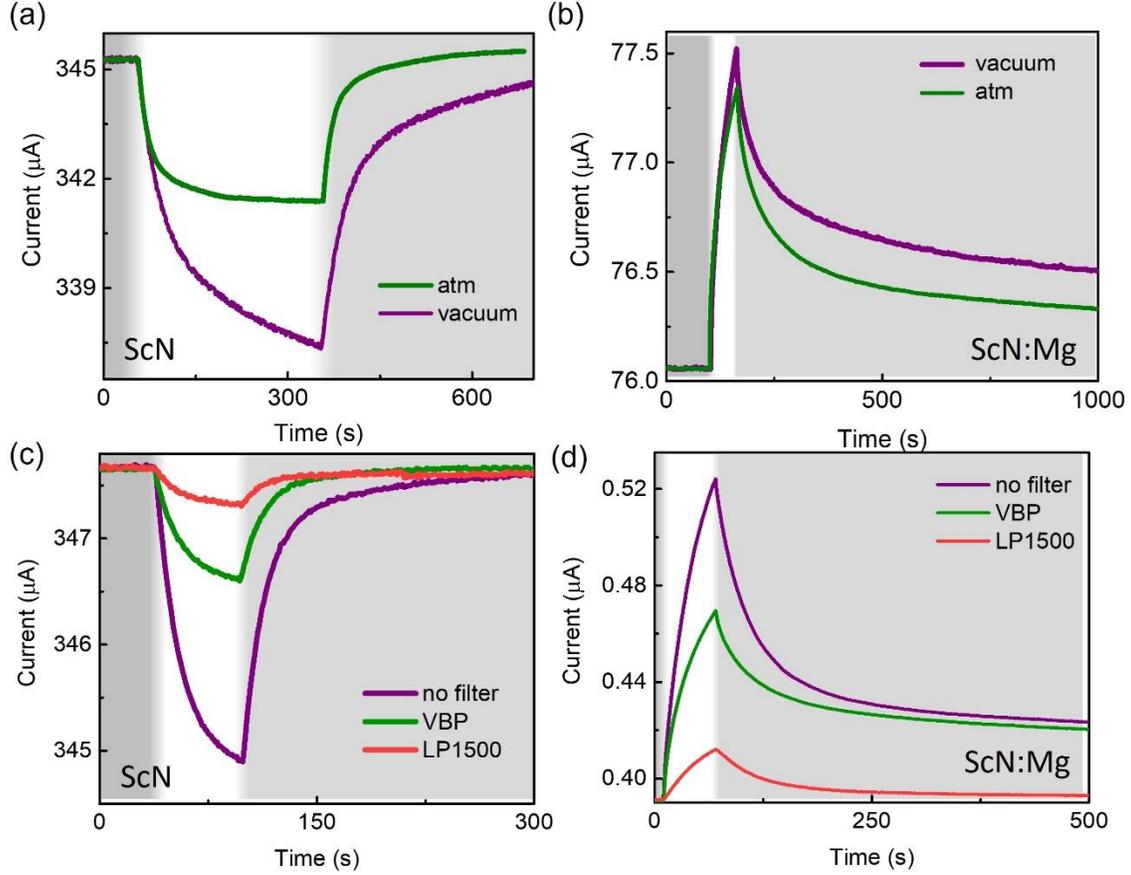

Figure S8: Photo response of (a) undoped ScN and (b) Mg-doped ScN measured in atmospheric and vacuum ( ~1×10$^{-5}$ torr) condition. Photo response using bandpass visible bandpass (VBP) filter and 1500 nm long pass (LP1500) filters of (c) as-deposited ScN and (d) Mg-doped ScN.

## 6. Synapse Power Consumption

The power consumption in a synapse is calculated in several ways[6]. Some researchers use equation 6 where energy consumed(dE) is the product of area of the device (S) , power of the optical pulse (P) having pulse width dt. This gives the energy of optical pulse incident on the device and does not include the material property of the synaptic device.

dE= S×P×dt   {6}

The more appropriate way to calculate energy consumption is by multiplying the read voltage (V)  photocurrent (I) and the duration of pulse (dt) as given in equation 7. However, comparing the energy consumption across devices of different dimension and stimuli width becomes difficult when calculated this way. In order to be able to compare the power consumption across different device dimension and pulse width, we normalize this equation and report the power density calculated as described in the main text.

$dE = V \times I \times dt$  {7}

The power density of the earlier demonstrated optoelectronic artificial synaptic devices along with those demonstrated here is listed in table S2. Only two terminal devices either with top-down or lateral contacts are listed. We see that the power density of ScN devices demonstrated here are in the range of other 2D-material and oxide material synaptic device and appreciably lesser than the only other III-nitride (AlN) based synaptic device.

Table S2: Power density comparison with some reported two-terminal artificial optoelectronic synapses.

| System | Power density nW/mm$^2$ | Contacts | |
|---|---|---|---|
| Nb:SrTiO$_3$ | 51 | top-down | [7] |
| ZnO/AlO | 22 | top-down | [8] |
| MoS$_2$ | 178 | top-down | [9] |
| Si Nanocrystals | 0.625 | lateral | [10] |
| CsPbBr$_3$ QD | 28 | lateral | [11] |
| AlN | $255 \times 10^9$ | lateral | [12] |
| ScN | 0.13 | lateral | This work |
| ScN:Mg | 0.65 | | |